\documentclass[]{jfm}
\usepackage{amsmath, amssymb, amsfonts, gensymb}
\usepackage{graphicx}
\usepackage[comma,authoryear]{natbib}
\graphicspath{ {Figures/} }

\newcommand{\diff}{\text{d}}

\newcommand{\Sc}{\text{Sc}}

\begin{document}
 
\title{Axisymmetric spreading of surfactant from a point source}
\author{Shreyas Mandre}
\affiliation{School of Engineering, Brown University, Providence RI 02912 USA}
\maketitle
\begin{abstract}
Guided by computation, we theoretically calculate the steady flow driven by Marangoni stress due to surfactant introduced on a fluid interface at a constant rate.
Two separate extreme cases, where the surfactant dynamics is dominated by the adsorbed phase or the dissolved phase are considered.
We focus on the case where the size of the surfactant source is much smaller than the size of the fluid domain, and the resulting Marangoni stress overwhelms viscous forces so that the flow is strongest in a boundary layer close to the interface.
We derive the resulting flow in a region much larger than the surfactant source but smaller than the domain size is described by approximating it with a self-similar profile.
The radially outward component of fluid velocity decays with the radial distance $r$ as $r^{-3/5}$ when the surfactant spreads in an adsorbed phase, and as $r^{-1}$ when it spreads in a dissolved phase.
Universal flow profiles that are independent of the system parameters emerge in both the cases.
Three hydrodynamic signatures are identified to distinguish between the two cases and verify the applicability of our analysis with successively stringent tests.
\end{abstract}

\section{Introduction}
Surfactant spreading on a liquid has received much attention for thin films \citep{Craster2009}, but not for deep layers of fluids.
Past attempts at analyzing Marangoni-stress driven flow on a deep layer of fluid have been in the context of thermo-capillary \citep*[][e.g.]{Bratukhin1967,Napolitano1979,Zebib1985,Carpenter1990} or thermo-soluto-capillary convection \citep{Bratukhin1968}, with the notable exception of \cite{Jensen1995} who analyzed transient dynamics from localized release of adsorbed surfactant. 
Interest has recently increased in the study of steady flow set by release of soluble amphiphiles at a constant rate \citep*{Roche2014,LeRoux2016}.
In this case, the surfactant is removed from the vicinity of the interface as it dissolves in fluid bath, establishing a state that changes very slowly with the surfactant concentration in the bath.
Consistent theoretical treatment of this type of Marangoni-driven surfactant advection is of fundamental interest and the topic of this article. 

% {\bf $<$Surfactant transport discussion here.$>$}
The general surfactant transport process coupled with the sorption kinetics and driven by a self-imposed Marangoni stress \citep*{Dukhin1995,Noskov1996,Eastoe2000,Young2009,Xu2013} is a problem with formidable complexity.
The surfactant concentration is governed by the equilibrium isotherm and the transient dynamics of adsorption and desorption.
The transport of surfactant is governed by the surfactant diffusion and advection by the flow established by the Marangoni stress.
The Marangoni stress itself depends on the relation between the instantaneous surfactant concentration at the interface and the surface tension. 
The flow that develops is governed by the Navier-Stokes equations, depending on the density and viscosity of the fluid, and any geometric parameters describing the fluid domain.
An analytical solution to the general coupled problem, while greatly desired, is not available.

For the case under consideration, a steady release of soluble amphiphilic surfactant through a point source on the interface has been reported to establish a quasi-steady flow near the surface of a deep pool \citep{Roche2014,LeRoux2016}.
The radial extent of this flow is finite for the surfactants studied \citep{LeRoux2016} and the axi-symmetric radial velocity profile on the surface of the fluid appears to be universal in shape \citep{Roche2014}.
In these studies, it was tacitly assumed that the surfactant spreads in a dissolved phase, and that the surface tension gradient driving the flow arises from the bulk concentration of the surfactant near the interface.
More recent experiments by \cite*{Mandre2017a} revealed a power-law decay of the surface velocity either with the distance $r$ from the surfactant source as $r^{-3/5}$ or as $r^{-1}$, and an accompanying depth-wise self-similar boundary layer profile.
Here we theoretically expound a possible reason for the two different exponents and the self-similar velocity profile in terms of two extremes in the surfactant dynamics.
In this article, we show that when the surfactant dynamics are dominated by the surface adsorbed phase, the fluid velocity decays as $r^{-3/5}$, and when dominated by the dissolved phase the decay is as $r^{-1}$.
The former corresponds to the surfactant dynamics in the Marangoni regime where the hydrodynamic time scale is so much faster than the sorption kinetics that little exchange of surfactant occurs between the surface and the bulk. 
The latter corresponds to the Gibbs regime with the sorption kinetics occuring much faster than the hydrodynamic time scale so that an equilibrium between surface and bulk surfactant may be assumed to have established.

Using numerical computations as initial guides, we develop here the mathematical solutions describing the axisymmetric flow resulting from a concentrated steady source of surfactant on the surface of a deep fluid layer.
We maintain the two-way coupling between the surfactant transport and the flow.
Instead of the more general problem of soluble surfactant dynamics, we consider the two extreme possibilities where the surfactant dynamics are dominated by the adsorbed phase (i.e. Marangoni regime) or the dissolved phase (i.e. Gibb's regime).
The specific mathematical models for the two cases of adsorbtion-dominated or dissolution-dominated surfactant transport are described in \S\ref{subsec:caseadsorbed}-\ref{subsec:casedissolved}.
Each case is simulated numerically (see \S \ref{subsec:numerical}) and the resulting flow in the region much larger than the surfactant source but much smaller than the flow domain is rationalized using similarity solutions (\S \ref{sec:insolsim} and \ref{sec:solsim}).
In the case of adsorption-dominated surfactant transport, we exploit a thin boundary layer structure of the flow near the fluid surface to make analytical progress.
The criteria for validity of the assumptions that underlie the boundary layer similarity solution for this case are also presented in \S\ref{subsec:validity}.
The similarity solution for dissolution-dominated surfactant transport is available due to \cite{Bratukhin1967}, which we specialize to the limit where the flow occurs in a boundary layer.
% (The similarity solution for the case of adsorbtion-dominated surfactant transport is our own result.)
We conclude in \S\ref{sec:conclusion} by presenting invariant hydrodynamic signatures that distinguish between the two extremes in surfactant dynamics.

\begin{figure}
\includegraphics{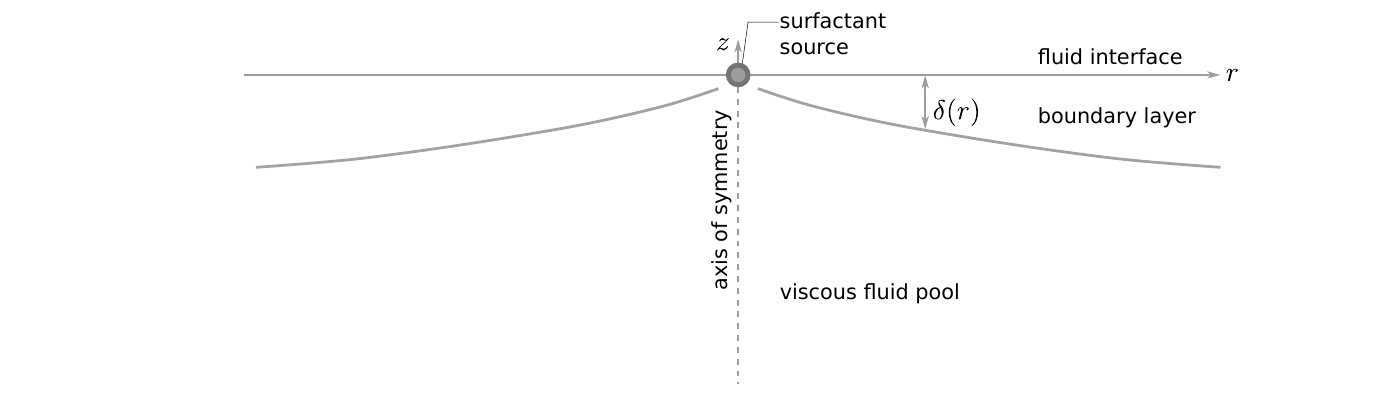}
\caption{Schematic setup of the problem. A point source located at the origin releases insoluble surfactant on the interface of a semi-infinite pool of fluid. The interface is along the $z=0$ plane. The Marangoni stress on the fluid interface arising from the non-uniform distribution of surfactant drives a flow in a boundary layer of thickness $\delta(r)$.}
\label{fig:Theory_Schematic}
\end{figure}

\section{Mathematical model}
Consider an semi-infinite bath of fluid with a free interface along the $z=0$ plane as schematically shown in Figure \ref{fig:Theory_Schematic}.
A surfactant is released at a constant rate from a point source located at the origin.
The axisymmetric forcing suggests description of the flow in cylindrical polar coordinates $(r,z)$ using the radial and axial components of velocity $u(r,z)$ and $w(r,z)$ respectively.

The fluid flow satisfies the incompressible Navier-Stokes equations
\begin{subequations}
\label{eqn:ns}
\begin{align}
 (ru)_r + r w_z &= 0, \label{eqn:rmom} \\
 (ru^2)_r + (rwu)_z  + rp_r &= \nu \left( ru_{zz} + (ru_r)_r -\dfrac{u}{r} \right), \label{eqn:zmom} \\
 (ruw)_r  + (rw^2)_z + rp_z &= \nu \left( rw_{zz} + (rw_r)_r \right), \label{eqn:mass}
% 
%  u_t + u u_r + w u_z &= -\dfrac{1}{\rho} p_r + \nu \left[ \dfrac{1}{r} \left( ru_r \right)_r -\dfrac{u}{r^2} + u_{zz} \right], \label{eqn:rmom}\\
%  w_t + u w_r + w w_z &= -\dfrac{1}{\rho} p_z + \nu \left[ \dfrac{1}{r} \left( rw_r \right)_r + w_{zz} \right], \label{eqn:zmom} \\
%  \dfrac{1}{r} (ru)_r + w_z &= 0. \label{eqn:mass}
\end{align}
\end{subequations}
where $p$ is the fluid pressure field divided by fluid density.
The fluid is quiescent far from the interface. 
The interface is assumed to be fixed at $z=0$, and the Marangoni stress arising from the non-uniform surface tension $\sigma$ implies
\begin{align}
 w=0 \quad \text{ and } \quad  \mu u_z = \sigma_r \quad \text{ at } \quad z=0. \label{eqn:nopenetration}
%  \label{eqn:marangoni}
\end{align}

\subsection{Case of adsorbtion-dominated surfactant dynamics}
\label{subsec:caseadsorbed}
\begin{figure}
\centerline{\includegraphics[width=0.93\textwidth]{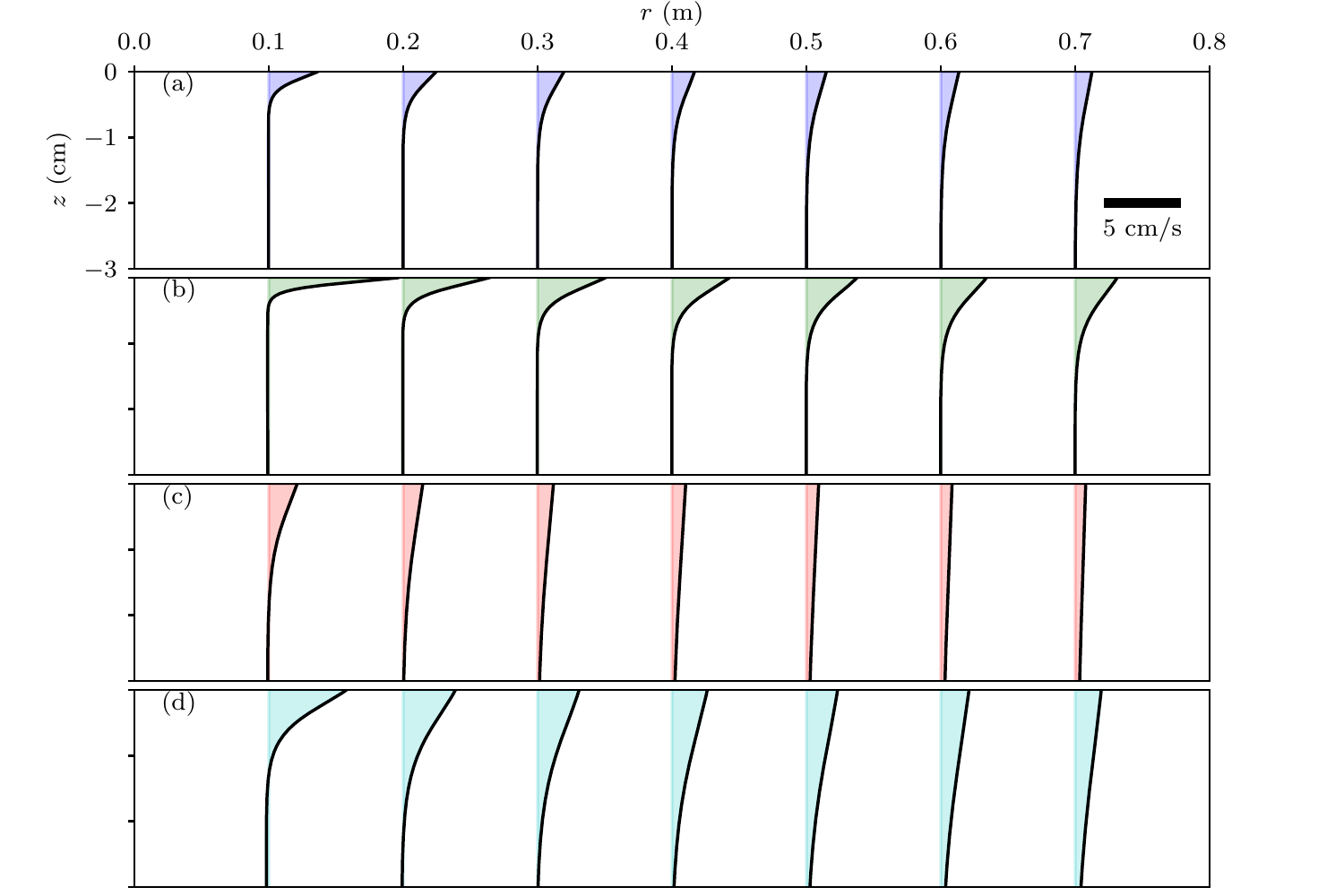}}
\caption{(Colour online) Sample radial velocity profiles near the free interface $z=0$, where the strongest flow occurs, obtained from the solutions of (\ref{eqn:ns}-\ref{eqn:marangonibc}). 
The profiles are plotted for 7 distances from the source corresponding to radial sections at $r$=0.1, 0.2, 0.3, 0.4, 0.5, 0.6, and 0.7 m as indicated by location of each curve. 
The four panels show: (a) $\nu=10^{-6}$ m$^2$/s, $K_2 = 10^{-2}$ m$^3$/s$^2$ (blue), (b) $\nu=10^{-6}$ m$^2$/s, $K_2 = 10^{-1}$ m$^3$/s$^2$ (green), (c) $\nu=10^{-5}$ m$^2$/s, $K_2 = 10^{-3}$ m$^3$/s$^2$ (red), (d) $\nu=10^{-5}$ m$^2$/s, $K_2 = 10^{-2}$ m$^3$/s$^2$ (cyan).}
\label{fig:bl_raw_panels}
\end{figure}

In case the sorption kinetics are not sufficiently fast relative to the hydrodynamic time scale, the surface tension of the interface is dominated by the dynamics of the adsorbed surfactant.
The surface tension $\sigma$ depends on the area concentration of the surfactant $c_2$, which we approximate for small concentrations to be linear as
\begin{align}
 \sigma = \sigma_0 - \Gamma_2 c_2,
 \label{eqn:insoleqnofstate}
\end{align}
where $\sigma_0$ is the interfacial tension without surfactant and $\Gamma_2$ is a material constant.
We assume the diffusion to be weak compared to advection, as is the case for most surfactants, so that diffusion of the surfactant can be neglected.
A quantitative criteria for the validity of this neglect is developed later in the article.
The surfactant is transported along the interface by advection implying
% \begin{subequations}
% \label{eqn:2dsurfactant}
\begin{align}
%  rc_{2,t} + (ruc_2)_r &= 0 \text{ at } z=0; \\
 2\pi ru(r,0) c_2 &= q_2 = \text{ constant},
\label{eqn:marangonibc}
\end{align}
% \end{subequations}
where $q_2$ is the strength of the point source.
Note that due to the linear relation between $c_2$ and $\sigma$ in \eqref{eqn:insoleqnofstate} and between $c_2$ and $q_2$ in \eqref{eqn:marangonibc}, the parameters $q_2$ and $\Gamma_2$ only influence the flow through the combination $K_2=\Gamma_2 q_2/2 \pi \mu$.
The only independent parameters in the problem are $K_2$ and $\nu$, both possessing purely kinematic dimensions.
The focus of this paper is on the steady flow that is established far behind the surfactant spreading front or as the surfactant dissolves in the bulk far away from the source and is depleted from the interface.
This condition is implemented computationally by introducing a surfactant sink on the outer boundary of the computational domain.

\subsection{Case of dissolution-dominated surfactant dynamics}
\label{subsec:casedissolved}
In the case of a soluble surfactant, the volumetric concentration $c_3$ dominates surfactant transport and governs the surface tension profile, which we represent as
\begin{align}
 \sigma = \sigma_0 - \Gamma_3 c_3 \qquad \text{at} \qquad z=0,
 \label{eqn:soleqnofstate}
\end{align}
where $\Gamma_3$ is a material parameter.
The surfactant, in this case, is transported within bulk of the fluid by an advection-diffusion process as
% \begin{subequations}
\begin{align}
%  c_{3,t} + u c_{3,r} + w c_{3,z} &= D\left[ \dfrac{1}{r} \left( r c_{3,r} \right)_r +  c_{3,zz} \right], \\
% \text{ and therefore at steady state } 
 u c_{3,r} + w c_{3,z} &= D \left[ \dfrac{1}{r} \left( r c_{3,r} \right)_r +  c_{3,zz} \right],
 \label{eqn:solmarangonibc}
\end{align}
% \end{subequations}
where $D$ is the diffusivity of the surfactant.
Note that the bulk diffusivity of the surfactant may not be neglected in this case, no matter how small its may be.
It is so because diffusion across the depth of the fluid layer governs the surface concentration, which in turn determines the Marangoni force.
Conservation of the surfactant is expressed in terms of the integrated flux of surfactant $q_3$ crossing a cylinder of radius $r$ as 
\begin{align}
 \int_{-\infty}^{0} \left[ 2\pi r u(r,z)  c_3(r,z)- D r c_{3,r}(r,z) \right]~\diff z = q_3=\text{constant} \qquad \text{for all } r.
 \label{eqn:solfluxcon}
\end{align}
Similar to the case of insoluble surfactant, the linear relations between $\sigma$, $c_3$ and $q_3$ imply that $\Gamma_3$ and $q_3$ only appear in the combination $K_3 = \Gamma_3 q_3/2\pi \mu$.
Thus $K_3$, $\nu$ and $D$ are the independent parameters describing the problem, all of them possessing purely kinematic dimensions.
The Schmidt number is defined as $\Sc = \nu/D$.
\begin{figure}
\centerline{\includegraphics[width=0.93\textwidth]{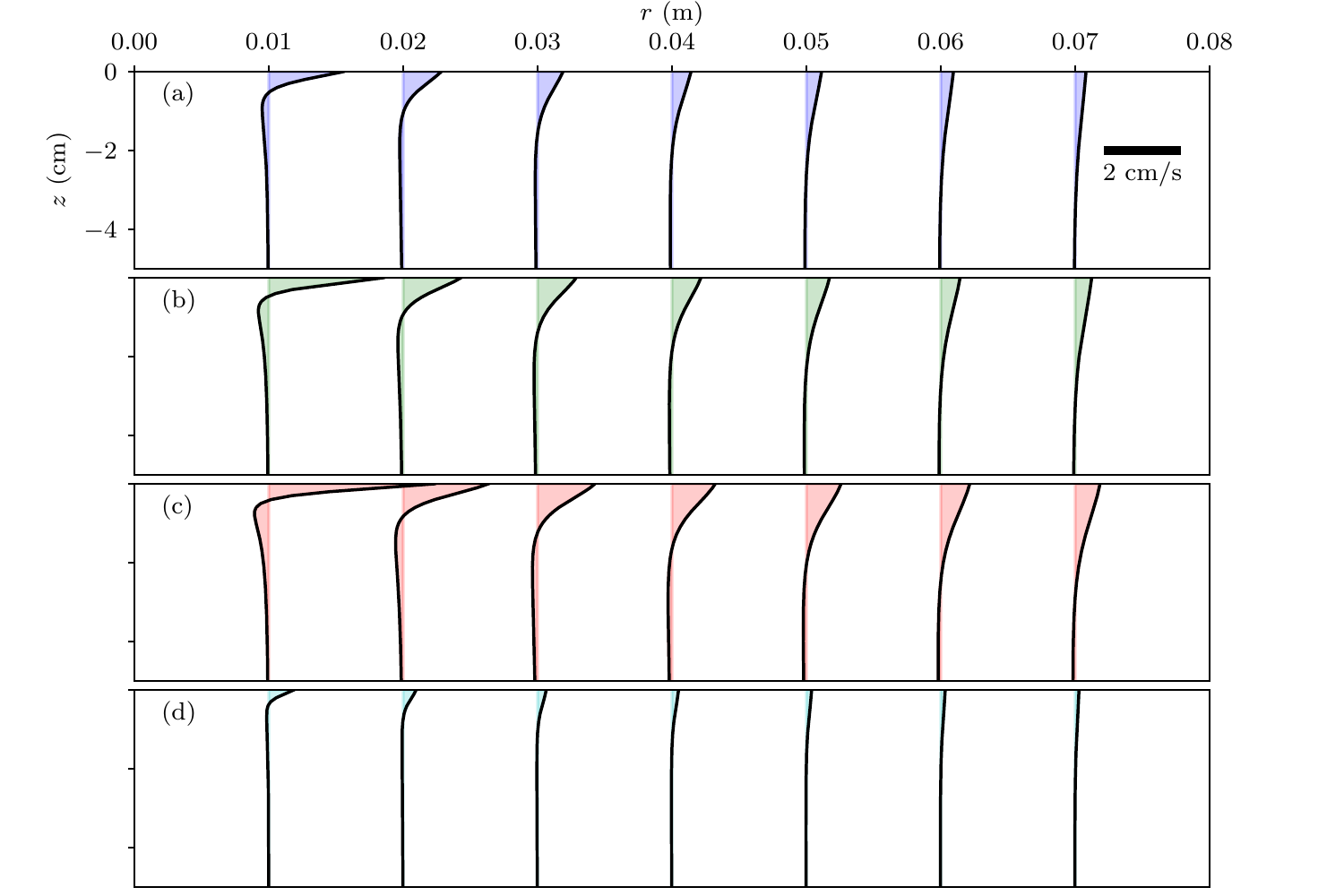}}
\caption{(Colour online) Sample radial velocity profiles near the free interface $z=0$ obtained from the solutions of (\ref{eqn:ns}), \eqref{eqn:soleqnofstate}, and \eqref{eqn:solmarangonibc}. 
The profiles are plotted for 7 distances from the source corresponding to radial sections from $r$=0.01 to 0.07 m as indicated by location of each curve. 
The four panels show: (a) $K_3 = 4\times 10^{-9}$ m$^4$/s$^2$ (blue), (b) $K_3 = 8 \times 10^{-9}$ m$^4$/s$^2$ (green), (c) $K_3 = 16 \times 10^{-9}$ m$^4$/s$^2$ (red), (d) $K_3 = 8 \times 10^{-9}$ m$^4$/s$^2$ (cyan). 
Here $\nu=2\times 10^{-5}$ m$^2$/s,  and $\Sc = 2$, except for panel (d), where $\nu=2 \times 10^{-6}$ m$^2$/s and $\Sc=0.2$.}
\label{fig:soluble_raw_panels}
\end{figure}

Although we have reduced the number of parameters in each case, they are still too numerous to furnish a useful non-dimensionalization.
For example, a length scale $\nu^2/K_2$ and a velocity scale $K_2/\nu$ may be constructed for the insoluble case, but as we find later, these parameter combinations do not represent the scales of length and velocity realized in the solution of the governing equations.
It is so because, by supposition, the region of interest for our analysis spans distances from the source much greater than $\nu^2/K_2$.
It is not {\it a priori} obvious how the length scale $\nu^2/K_2$ and the distance from the source must be combined to derive the appropriate length scale for the Marangoni flow (see Napoitano, 1979, \nocite{Napolitano1979} for a more detailed dimensional analysis).
Therefore, we first attempt a computational solution of the dimensional governing equations, and upon examining their structure construct a suitable non-dimensionalization.

\subsection{Numerical solutions}
\label{subsec:numerical}
Transient versions of the governing equations were implemented and solved using COMSOL's Computational Fluid Dynamics and Mathematical Modeling module starting from a static fluid layer with a clean interface until a steady state is reached.
The domain was chosen to be a cylinder large enough (radius 2.5 m and depth 2.5 m for the insoluble case, and radius 0.48 m and depth 0.48 m for the soluble case) to approximate an infinite domain, but small enough so that for the parameters chosen no flow instabilities appeared. 
The $z<0$ region of the cylinder is filled with a fluid possessing density in the range $10^{3}-10^4$ kg/m$^3$ and dynamic viscosity $\mu$ ranging from $10^{-1}-10^{-3}$ Pa s. 
The point source of surfactant located at the origin is assigned a small finite extent (radial extent $r_0 \approx 10^{-3}$ m) so that the singularity at $r=0$ is regularized.
Outflow (i.e. zero gage pressure) boundary condition are imposed at the outer radius and lower boundary to better approximate an infinite domain.
Imposing other boundary conditions such as no-slip or slip without penetration does not significantly affect the flow in the region much larger than $r_0$ but smaller than the domain size.
To achieve a steady state, the surfactant is absorbed on the outer boundary of the cylinder by imposition of $c_{2,3}=0$, while a no-flux condition applies everywhere else on the boundary.
We used an non-uniform unstructured triangular mesh and a non-uniform structured rectangular mesh for discretization.
The discretization is finer near the surfactant source and the interface to resolve the presence of the boundary layer.
The grid was successively refined to test for numerical convergence and confirm the self-similar flow structure.

Sample steady state radial flows for four cases are shown in Figure~\ref{fig:bl_raw_panels} and \ref{fig:soluble_raw_panels}.
The flow is fastest at the interface, and falls off rapidly with depth over a length of O(1 cm) indicating a boundary layer structure driven by Marangoni stresses.
% Comparison between Figure \ref{fig:bl_raw_panels}(a-b) and (c-d) shows that the surface flow increases and the apparent boundary layer thickness decreases with increasing insoluble surfactant source strength, $K_2$.
% Similarly, Figure \ref{fig:soluble_raw_panels}(a-c) shows such behaviour for soluble case with parameter $K_3$.
% In both cases, the surface flow decreases and the apparent boundary layer thickness increases with increasing kinematic viscosity $\nu$, as can be seen by by comparing panels (a,c) with (b,d) of Figure~\ref{fig:bl_raw_panels}, and panels (b) with (d) of Figure~\ref{fig:soluble_raw_panels}.
The dependence of the velocity and length scales on the problem parameters will be explained using a boundary layer approximation and self-similar solution of the governing equations.

\begin{figure}
\centerline{\includegraphics[width=0.7\textwidth]{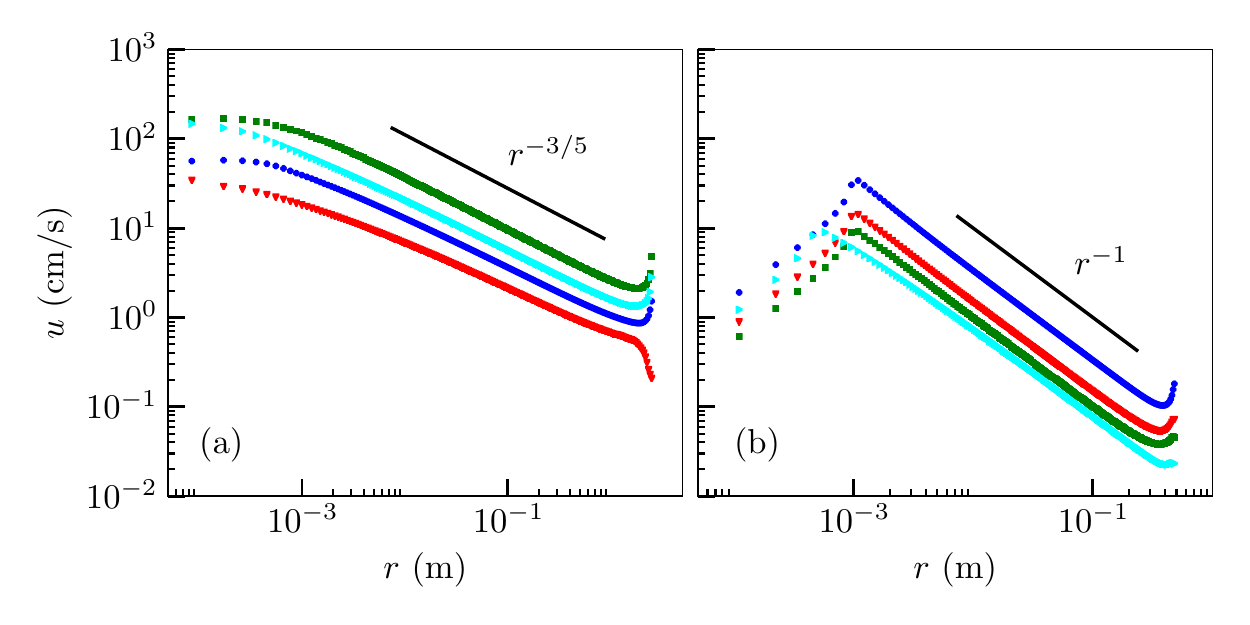}}
\caption{(Colour online) Surface radial velocity as a function of radial distance. (a)~The four cases of insoluble surfactant shown in Figure~\ref{fig:bl_raw_panels} with identical colour code. Solid black line shown the power law $r^{-3/5}$. (b)~The four cases of soluble surfactant shown in Figure~\ref{fig:soluble_raw_panels} with identical colour code. Solid black line shown the power law $r^{-1}$.}
\label{fig:PowerLaws}
\end{figure}

A simple description of the flow in terms of power law decay away from the surfactant source may be derived using dominant balances as follows.
At a radius $r$ from the source, let us assume that the radial velocity decays as a power law denoted by $u(r)$ and the surfactant concentration as $c_{2,3}(r)$.
A balance between inertia (per unit mass) which scales as $u^2/r$ and viscous forces which scale as $\nu u/\delta^2$, where $\delta(r)$ is the expected boundary layer thickness, implies $\delta = (\nu r/ u)^{1/2}$.
When the surfactant spreads in an adsorbed phase, its conservation implies $2 \pi r u c_2 = q_2$.
Combining this with a balance between the Marangoni stress $\Gamma_2 c_2/r$ and the shear stress on the interface $\mu u/\delta$, yields 
\begin{align}
u \propto \dfrac{K_2^{2/5} \nu^{1/5}}{r^{3/5}}, \quad \delta \propto \dfrac{\nu^{2/5} r^{4/5}}{K_2^{1/5}}, \text{ and } \dfrac{c_2}{q_2} \propto \dfrac{1}{K_2^{2/5} \nu^{1/5} r^{2/5}}.
\label{eqn:insolpowerlaws}
\end{align}

% On the other hand, surfactant conservation in the dissolved phase implies $u c_3 r \delta \Sc^{1/2} = q_3$ (here we have assumed advection to dominate over diffusion of surfactant).
% The Marangoni stress balance $\Gamma_3 c_3/r$ with shear stress $\mu u/\delta$ yields
% \begin{align} 
% u \propto \dfrac{K_3^{1/2}}{r}, \quad \delta \propto \dfrac{\nu^{1/2} r}{K_3^{1/4}}, \text{ and } \dfrac{c_3}{q_3} \propto \dfrac{1}{K_3^{1/4} \nu^{1/2} r}. 
% \label{eqn:solpowerlaws}
% \end{align}
A similar scaling analysis for the case of dissolved surfactant may be used to rationalize the $r^{-1}$ scaling of velocity but not the dimensional pre-factor. 
It can readily be seen from momentum conservation that if $u\propto r^n$, then $\delta \propto r^{(1-n)/2}$.
Surfactant transport occurs in a layer of thickness $\delta \Sc^{-1/2} \propto r^{(1-n)/2}$.
The Marangoni stress at the surface scales as $u/\delta \propto r^{(3n-1)/2}$, which implies the surfactant concentration scales as $c_3 \propto r^{(3n+1)/2}$.
Therefore, the surfactant flux is $q_3 \propto ruc_3 \delta \propto r^{2(1+n)}$. 
Since $q_3$ is independent of $r$, we conclude $n=-1$, implying $u \propto r^{-1}$, $\delta \propto r$, and $c_3 \propto r^{-1}$.
The reason for the failure of this simple scaling argument to yield the dimensional pre-factors will be clarified later.

In Figure~\ref{fig:PowerLaws}, we plot the numerically obtained radial velocity on the surface as a function of $r$ and compare it with the afore-derived power laws.
(The kink in the profiles at $r=10^{-3}$ m for the dissolution-dominated case corresponds to the abrupt jump in the surfactant flux from a non-zero value inside the source to zero outside.)
The power-law behavior compares well with the numerical solution in an intermediate range of radial distances, which are much larger than the source but smaller than the domain size.

\section{Self-similar profiles}
The power-law description paves way for a self-similar description of the flow in this intermediate region, which we derive next.
The simple balances presented above, including the dimensional pre-factors for the adsorbed surfactant case, are closely associated with a self-similar flow with a universal velocity and surfactant profile, which is the topic of the rest of the article.

\subsection{Adsorption-dominated surfactant transport}
\label{sec:insolsim}

\begin{figure}
\centerline{\includegraphics[width=11.5cm]{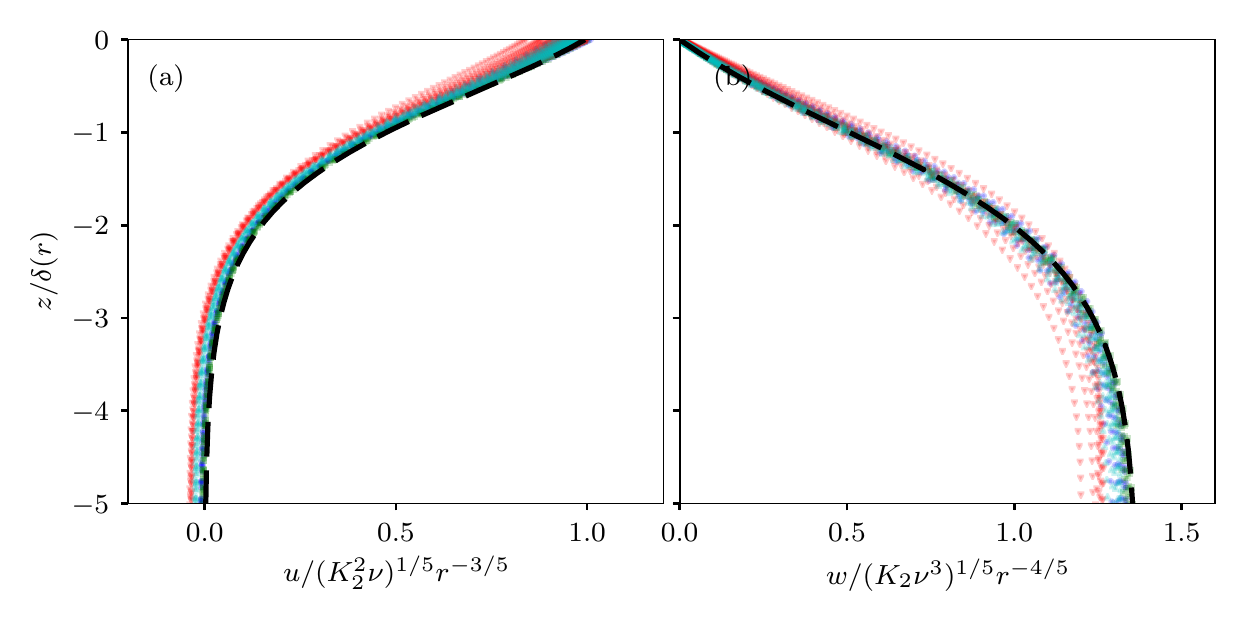}}
\caption{(Colour online) Comparison of the computational solutions with the similarity solution in the boundary layer for the case of adsorbed surfactant transport. 
(a) The radial component and (b) the axial component of fluid velocity, scaled according to the similarity solution predictions, is plotted (green symbols) for all the profiles shown in Figure \ref{fig:bl_raw_panels}. 
The solution of (\ref{eqn:selfsimilarode}-\ref{eqn:selfsimilarbc}) obtained using shooting method (dashed curve) is also shown for comparison. 
% The solid line shows $f'(\xi)$ derived as a solution of (\ref{eqn:selfsimilarode}-\ref{eqn:selfsimilarbc}) for comparison.} 
}
\label{fig:bl_scaled}
\end{figure}

% Now we derive similarity solution for the cases where the surfactant is adsorbed to the interface and compare it with the results of the numerical solutions.
The power law in the surface velocity profile implies scale free dynamics, and therefore a self-similar flow.
The self-similar flow can be described using a similarity variable
\begin{align}
 \xi = \dfrac{z}{\delta(r)}, \text{ where } \delta = \dfrac{\nu^{2/5} r^{4/5}}{K_2^{1/5}}
 \label{eqn:bldef}
\end{align}
is the boundary layer thickness according to (\ref{eqn:insolpowerlaws}). Assuming $\delta \ll r$, the governing fluid equations simplify to the Prandtl boundary layer equations
\begin{align}
 uu_r + w u_z = \nu u_{zz}, \quad p_z = 0, \quad (ur)_r + rw_z = 0, \label{eqn:prandtlbleqn}
\end{align}
in a region of length $O(\delta)$ near the interface.
Outside this region, the flow is weak, and therefore $u$, $w$ and $p$ are neglected there.
A separate treatment of the outer region follows later. 

Continuity \eqref{eqn:mass} may be satisfied by using a self similar form for the velocity components with the dimensional pre-factor determined from \eqref{eqn:insolpowerlaws} as 
% Stream function $\psi(r,z)$ can be used to satisfy mass conservation  by defining
% \begin{align}
% r u = \psi_z, \quad r w = -\psi_r, \quad  \psi(r,z) = {K_2^{2/5} \nu^{1/5}}{r^{2/5}} \delta(r) f(\xi),
% \end{align}
% The following relations for the derivatives of these quantities are handy:
% \begin{align}
%  \dfrac{\partial \xi}{\partial z} = \dfrac{1}{\delta(r)}, \quad \text{ and }  \quad \dfrac{\partial \xi}{\partial r} = -\dfrac{\xi}{r} \left( \dfrac{r \delta'(r)}{\delta(r)} \right) = -\dfrac{4 \xi}{5r}. 
% \end{align}
% According to this ansatz, 
\begin{align}
 u(r,z) = \dfrac{K_2^{2/5} \nu^{1/5}}{r^{3/5}} f'(\xi), \quad \text{ and } \quad  
 w(r,z) =-\dfrac{{K_2^{1/5} \nu^{3/5}}}{5r^{4/5}}  \left( 6 f(\xi) - {4\xi} f'(\xi)\right). \label{eqn:ansatz}
\end{align}
where $f$ is function to be determined. 

Substituting this ansatz in \eqref{eqn:prandtlbleqn} yields
\begin{align}
f'''(\xi) + \dfrac{3}{5} f'(\xi)^2  + \dfrac{6}{5} f(\xi)f''(\xi) = 0. \label{eqn:selfsimilarode}
\end{align}
\begin{figure}
% \vspace{5cm}
\centerline{\includegraphics[width=11.5cm]{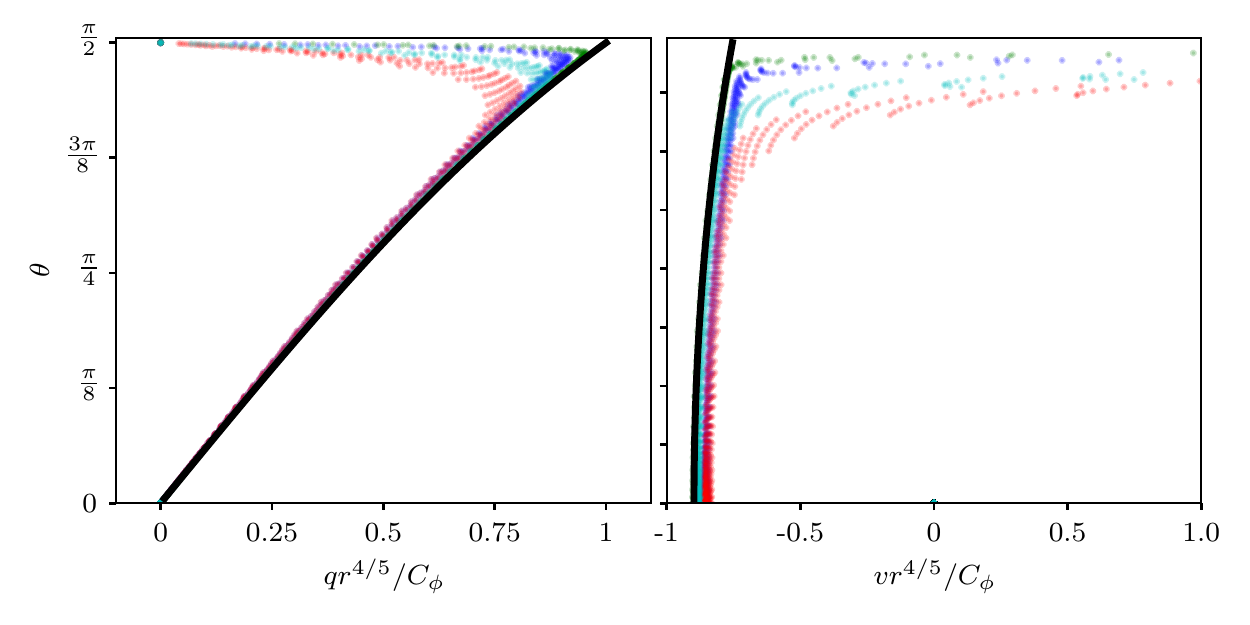}}
\caption{(Colour online) Comparison of the computational solutions with the similarity solution in the outer region for the case of adsorbed surfactant transport.
(a) Azimuthal velocity component and (b) the radial velocity component, in the range 0.02 m$<\rho<$0.1 m obtained from computational solutions (green symbols) is the scaled according to and compared with the analytical expression in equation \ref{eqn:outersol} (solid black curves).}
\label{fig:outer_scaled}
\end{figure}

% \subsection{Case of Insoluble surfactant}
% The boundary conditions for the case of the insoluble surfactant may be written as
% \begin{align}
%  ru(r,z=0) c_2(r) &= q_2 = \text{constant}, \\
%  \mu u_z(r, z=0) &= - \Gamma c_{2,r}.
% \end{align}
% These equations represent the surfactant, and the Marangoni stress balance.
The boundary conditions \eqref{eqn:nopenetration}, \eqref{eqn:insoleqnofstate} and \eqref{eqn:marangonibc} can be combined into one by eliminating $c_2(r)$,to get
\begin{align}
u_z(r, z=0) = -\dfrac{\Gamma_2 q_2}{2\pi \mu} \left( \dfrac{1}{r u(r, 0)} \right)_r = -K_2 \left( \dfrac{1}{r u(r, 0)} \right)_r. \label{eqn:insolselfsimilarbc}
\end{align}
% which shows that $\Gamma_2$, $q_2$ and $\mu$ always appear in the combination $K_2$.

% In order to substitute the ansatz in the boundary condition, we evaluate
% \begin{align}
%  \left( \dfrac{1}{r u(r, z)} \right)_r = \left( \dfrac{1}{a \Psi(r) f'(\xi)} \right)_r = -\left( \dfrac{1}{r a \Psi(r) f'(\xi)^2} \right) \left( n f'(\xi) - {m\xi} f''(\xi) \right)
% \end{align}
% which 
Upon substitution of the ansatz \eqref{eqn:ansatz} in \eqref{eqn:insolselfsimilarbc}, the condition \eqref{eqn:nopenetration} and the stagnation of the fluid outside the boundary layer yields a full set of boundary conditions for $f$ as
\begin{align}
f(0) = 0, \quad f''(0) f'(0) = \dfrac{2}{5}, \quad \text{and} \quad f'(-\infty) = 0. \label{eqn:selfsimilarbc}
\end{align}
The simple dominant balance, which we used to derive the power-law expressions in \eqref{eqn:insolpowerlaws}, can be seen throughout this derivation as the balance between coefficients of the terms representing the respective physical effects.
In this manner, the simple dominant balance analysis represents the more detailed derivation based on self-similarity.

\begin{figure}
\centerline{\includegraphics[width=11.5cm]{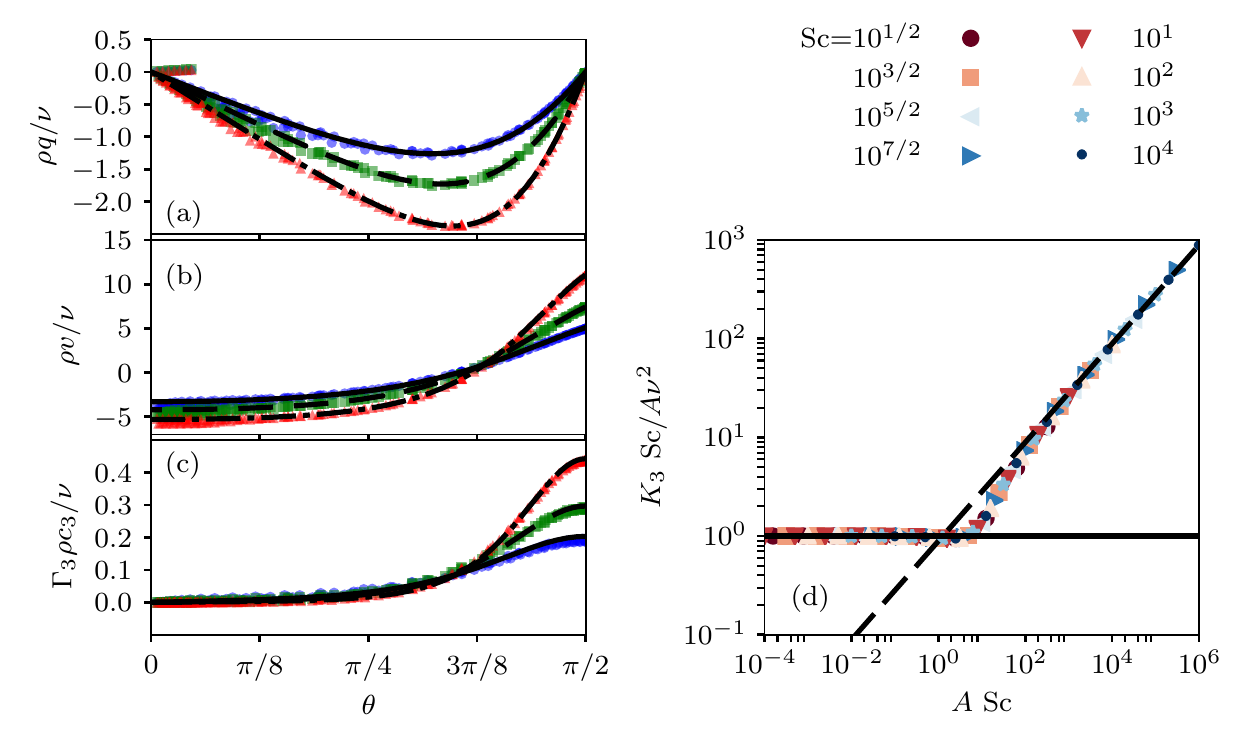}}
\caption{(Colour online) Similarity solution for the case of soluble surfactant and comparison with rescaled computational results from Figure \ref{fig:soluble_raw_panels}(a-c) with identical colour code.
(a) Azimuthal velocity, (b) radial velocity, and (c) surfactant concentration. 
Symbols indicate results of rescaled computational solutions $K_3 \Sc/\nu^2=20$ (blue circles), 40 (green squares), and 80 (red triangles) corresponding to panels in Figure \ref{fig:soluble_raw_panels} (a-c) respectively. 
Black curves indicate similarity solutions corresponding to $K_3 \Sc/\nu^2=20$ (solid), 40 (dashed), and 80 (dot-dashed).
(d) Relation between $A$ and the strength of Marangoni stress quantified by a single parameter $K_3 \Sc /\nu^2A$ from \eqref{eqn:A} for different $\Sc$ shown in the legend. 
The solid line shows the asymptotic value $K_3 \Sc /\nu^2A=1$ for $A\ll 1$ and the dashed line shows  $K_3 \Sc /\nu^2A=(\pi \Sc A)^{1/2}/2$ for $A \gg 1$.
}
\label{fig:sol_bl_scaled_bratukhin}
\end{figure}
% \begin{align}
%  \dfrac{a \Psi(r)}{r \delta(r)} f''(0) = K_2 n \left( \dfrac{1}{r a \Psi(r) f'(0)} \right).
% \end{align}
% Substituting $\delta(r)$ and equating the dimensional, the power law, and the self-similar variables separately yields
% \begin{align}
% a = K_2^{2/5} \nu^{1/5}, \qquad \Psi(r) = r^{2/5}, \qquad f''(0) f'(0) = n = \dfrac{2}{5}.
% \end{align}

% \vspace{3mm}
% The solution so far may be summarized in terms of a self-similar profile $f(\xi)$ as:
% \begin{align}
% \begin{split}
%  \psi(r,z) = (a \nu)^{1/2} r^{6/5} f(\xi), \text{ where } a=K_2^{2/5} \nu^{1/5}, \quad \xi = \dfrac{z}{\delta(r)}, \quad \delta(r) = \left( \dfrac{\nu }{a }\right)^{1/2} r^{4/5}, \\
%  u(r,z) = ar^{-3/5} f'(\xi), \quad w(r,z) = -\dfrac{ (\nu a)^{1/2} }{5} r^{-4/5}  \left( 6f(\xi) - 4\xi f'(\xi)\right), \quad c(r) = \dfrac{q}{a r^{2/5} f'(0)}, \\
%  n=\dfrac{2}{5}, \quad m = \dfrac{4}{5}.
% \end{split}
% \end{align}
% 
% The ordinary differential equation for $f(\xi)$
% \begin{align}
%  f'''(\xi) +  \dfrac{3}{5} f'(\xi)^2  +  \dfrac{6}{5} f(\xi)f''(\xi) = 0,
%  % - \dfrac{8}{5} \xi f''(\xi) f'(\xi) 
% \label{eqn:selfsimilarode}
% \end{align}
% is to be solved with the boundary conditions
% \begin{align}
%  f(0) = 0, \qquad f''(0) f'(0) = \dfrac{2}{5}, \qquad f'(-\infty) = 0.
% \label{eqn:selfsimilarbc}
% \end{align}

Equations \eqref{eqn:selfsimilarode} and \eqref{eqn:selfsimilarbc} are numerically solved using a shooting method, which we outline next. 
The method starts with a guess for $f''(0)$ and sets the corresponding $f'(0)$ using \eqref{eqn:selfsimilarbc}.
The initial value problem with the guessed initial condition is then solved numerically using a fourth order Runge Kutta method and its asymptotic behavior as $\xi \to -\infty$ is examined. The solution asymptotically either diverges to $\infty$ or $-\infty$ as $f\sim C \xi^{2/3}$, thereby violating the far-field boundary condition in \eqref{eqn:selfsimilarbc}. 
However, between the cases that diverge to $\infty$ and those that diverge to $-\infty$ is one solution that remains bounded.
For this case $f$ approaches a constant value, $f_\infty$.
This possibility may be examined by making the ansatz $f = f_\infty + \epsilon g(\xi) + \dots$ for $\epsilon \ll 1$, resulting in
\begin{align}
 f(\xi) = f_\infty + \epsilon \left( a_1 + b_1 \xi + c_1 e^{-6f_\infty \xi/5}  \right) + O(\epsilon^2),
\end{align}
where $a_1$, $b_1$ and $c_1$ represent arbitrary constants of integration.
Note that $b_1$ must be zero so that asymptotic ordering of the solution is maintained as $\xi \to -\infty$ and $a_1$ may be absorbed into $f_\infty$.
The objective of the shooting method is to guess the initial value $f''(0)$ such that this bounded solution with $b_1=0$ is asymptotically achieved.
We find this solution by successive bisection of the interval of $f''(0)$ with end points that lead to diverging solutions with opposite signs.
This bisection was implemented manually to determine that $f''(0) = 0.402287361293201$ solves (\ref{eqn:selfsimilarode}-\ref{eqn:selfsimilarbc}) numerically to 15 digit accuracy.
Correspondingly, $f'(0) = 0.994314110973191$ and $f_\infty = -1.13886447085041$, which leads to $u(r,0) = f'(0) K_2^{2/5} \nu^{1/5} r^{-3/5}$.
The axial velocity just outside the boundary layer is $w(r, \xi\to -\infty) = -1.2f_\infty K_2^{1/5} \nu^{3/5}r^{-4/5}$.
The resulting self-similar profile for $u$ and $w$ based on the solution for $f(\xi)$ and its derivatives is shown in Figure \ref{fig:bl_scaled}.

The similarity solution is compared with the four direct computational solution in Figure \ref{fig:bl_scaled}.
The reduction of the variability between the profiles resulting from the differences in $K_2$, $\nu$, and $r$, and the excellent comparison with the $f(\xi)$ and its derivatives computed as a solution of \eqref{eqn:selfsimilarode} and \eqref{eqn:selfsimilarbc}, verifies the validity of the similarity solution.

Outside the boundary layer, the flow retains its self-similar structure, but with a different scaling.
The viscous stresses may be neglected to leading order, and a potential flow driven by volume flux of fluid entraining the boundary layer may be used to describe the resulting flow.
The velocity potential, denoted $\phi(\rho,\theta)$ in terms of spherical polar coordinates $(\rho, \theta)$ defined by $r = \rho \sin \theta$ and $z=-\rho \cos\theta$, with the corresponding velocity components $v=\phi_\rho$ and $q=\phi_\theta/\rho$ respectively, satisfies
% with $u=\phi_r$ and $w=\phi_z$, incompressibility implies 
\begin{align}
 \nabla^2 \phi = \dfrac{1}{\rho^2} \left(\rho^2 \phi_\rho \right)_\rho + \dfrac{1}{\rho^2\sin\theta}\left(\sin\theta \phi_{\theta}\right)_\theta = 0,
\end{align}
with the matching condition of the fluid flux into the boundary layer,
\begin{align}
 \dfrac{1}{\rho}\phi_\theta(\rho, \theta\to \dfrac{\pi}{2}^{-}) =  w(\rho, \xi\to -\infty) = -1.2f_\infty K_2^{1/5} \nu^{3/5}\rho^{-4/5}.
 \label{eqn:matchingvel}
\end{align}
A standard solution of Laplace equation in the form $\phi = C_\phi r^n f_o(\cos\theta)$ may be sought.
Substituting this form in the boundary condition yields $n=1/5$, and the Legendre differential equation for $f_o$ as
\begin{align}
 (1-s^2) f_o''(s) - 2 s f_o'(s) + n(n+1) f_o = 0.
\end{align}
A solution may be written in terms of the regular Legendre function $P_\frac{1}{5}$, which solves this differential equation, as
\begin{align}
\begin{split}
 &\phi = C_\phi r^{1/5} P_\frac{1}{5}(\cos\theta), \quad 
 v = \dfrac{1}{5} C_\phi r^{-4/5} P_\frac{1}{5}(\cos\theta), \quad
 q = -C_\phi r^{-4/5} P'_\frac{1}{5}(\cos\theta) \sin\theta, \\
 &\text{where } C_\phi = \dfrac{6 f_\infty K_2^{1/5} \nu^{3/5}}{5P_\frac{1}{5}'(0)}.
\end{split}
\label{eqn:outersol}
\end{align}
(Note that $C_\phi$ here has SI units of m$^{9/5}$/s.)

This solution is compared with the computational solution in Figure \ref{fig:outer_scaled}.
To isolate the intermediate range of scales much larger than the source and the boundary layer thickness, but much smaller than the computational domain, this comparison is limited to computational points with 0.02 m $< \rho <$ 0.1 m.
The raw values of $v$ and $q$ in this region vary by about two orders of magnitude for a fixed $\theta$, however upon accounting for the dimensional factor $C_\phi$ and the spatial scaling factor of $\rho^{-4/5}$ the variation reduces to within 10\%.

The momentum balance in the outer region reduces to the Bernoulli equation, and yields the pressure variation as $p = -\dfrac{1}{2} ( v^2 + q^2)$.
Using this relation, the pressure may be estimated to leading order to be 
\begin{align}
 p = -\dfrac{C_\phi^2 }{2\rho^{8/5}} \left( \dfrac{P^2_{\frac{1}{5}} (\cos\theta) }{25} + P'^2_{\frac{1}{5}} (\cos\theta) \sin^2\theta \right).
 \label{eqn:outerpressure}
\end{align}
The pressure variation across the boundary layer is negligible due to its thinness, and therefore the pressure everywhere in the fluid is approximated by \eqref{eqn:outerpressure}.

\subsection{Asymptotic criteria for the validity of self-similar solution}
\label{subsec:validity}
We now revisit the assumptions made in deriving the similarity solution, and where possible derive the quantitative criteria that for their validity.
Perhaps the most striking assumption underlying this analysis is simplification of the surfactant transport processes. 
Since it is the premise of our approach that the velocity field encodes information about the surfactant transport, we do not examine the validity of the assumptions related to surfactant transport.
Instead, in this article we merely pursue the logical conclusion of these assumptions, so that they can be compared against experimental measurements. 
However, many other assumptions were made in the interest of semi-analytical results, and next we verify their validity.

\subsubsection{Boundary layer approximation}
The application of the boundary layer approximation implies $\partial_r \ll \partial_z$ and that the outer (radial) flow is much weaker than the one in the boundary layer.
The former implies $r \gg \delta$, or equivalently from \eqref{eqn:bldef}
\begin{align}
\dfrac{\nu^2}{K_2} \ll r.
%  r \gg \dfrac{\nu^{2/5} r^{4/5}}{K_2^{1/5}}. \\
% 1 \gg \dfrac{\nu^{2/5} r^{1/5}}{K_2^{1/5}}.
\label{eqn:blapproxcriteria}
\end{align}
Furthermore, the neglect of the outer radial velocity, which scales as $C_\phi/r^{4/5}$ from \eqref{eqn:outersol}, compared to the boundary layer radial velocity, which scales as ${K_2^{2/5} \nu^{1/5}}/{r^{3/5}}$ from \eqref{eqn:ansatz}, is equivalent to \eqref{eqn:blapproxcriteria}.
The criterion \eqref{eqn:blapproxcriteria} implies the similarity solution to be valid at radial distances much larger than the length scale that can be constructed from dimensional analysis.
% \begin{align}
% \dfrac{K_2^{1/5} \nu^{3/5}}{r^{4/5}} \ll \dfrac{K_2^{2/5} \nu^{1/5}}{r^{3/5}}. \\
% \dfrac{K_2^{1} \nu^{3}}{r^{4}} \ll \dfrac{K_2^{2} \nu^{1}}{r^{3}}. \\
% \dfrac{\nu^{2}}{r} \ll {K_2^{1} } \\
% 
% \end{align}

\subsubsection{Neglect of surfactant diffusion}
According to \eqref{eqn:marangonibc}, the surfactant concentration profile also decays as a power law, implying that the radial length scale for variation in $c_2$ is $r$.
The domination of the advective flux $uc_2$, which scales as ${K_2^{2/5} \nu^{1/5}}/{r^{3/5}} \times c_2$ from \eqref{eqn:ansatz}, over the diffusive flux, which scales as $D_2/r \times c_2$, $D_2$ being the diffusivity of the surfactant on the surface, implies
\begin{align}
% {K_2^{2/5} \nu^{1/5}}/{r^{3/5}} \gg D_2/r \\
% {K_2^{2} \nu^{1}}/{r^{3}} \gg D_2^5/r^5
r \gg \sqrt{\dfrac{D_2^5}{K_2^2 \nu}}.
\end{align}
% Note that this comparison essentially identifies the Peclet number for the surfactant transport. 

\subsubsection{Flatness of the interface}

We had neglected the interface deformation motivated by the observations of \cite{Roche2014} and \cite{Mandre2017a}, who report no perceptible deformation of the interface was observed, except for a liquid bridge conecting the interface to the conduit conveying the surfactant solution.
The closed form similarity solution allows us to quantitatively estimate the deformation of the interface due to the non-uniform pressure at the interface.
Noting that $P_\frac{1}{5}(0) = \pi^{1/2}/\Gamma(11/10) \Gamma(2/5)$ and $P'_\frac{1}{5}(0) = -2 \pi^{1/2}/\Gamma(3/5)\Gamma(-1/10)$ \citep[see][\S 8.6]{Abramowitz1964}, where $\Gamma$ is the Gamma function, the interface pressure may be approximated as
% \begin{align}
%  p = -\dfrac{\pi C_\phi^2 }{2r^{8/5}} \left( \dfrac{1}{25 \Gamma(11/10)^2 \Gamma(2/5)^2}+ \dfrac{4}{\Gamma(3/5)^2\Gamma(-1/10)^2} \right).
% \end{align}
\begin{align}
 p = -  C_p \dfrac{K_2^{2/5} \nu^{6/5}}{ r^{8/5}} \quad \text{where} \quad C_p = \dfrac{18 f^2_\infty}{25} \left( \dfrac{P^2_{\frac{1}{5}} (0) }{25 P'^2_\frac{1}{5}(0)} + 1 \right) \approx 0.955092285086512.
\end{align}
%where $C_p = 0.95509228508651167$ is the dimensionless constant.

To leading order of the flat-interface approximation, a combination of gravity and surface tension maintain the interface flat against this pressure. 
The change in elevation of the interface, $H(r)$, caused by the fluid pressure is given by
\begin{align}
% \sigma \dfrac{1}{r} \left( r H_r \right)_r - \rho g H = \rho p. \\
\dfrac{1}{r} \left( r H_r \right)_r - \dfrac{H}{l_c^2} = \dfrac{p}{\sigma} = -  C_p \dfrac{K_2^{2/5} \nu^{6/5}}{ \sigma r^{8/5}}.
\end{align}
where $l_c$ is the capillary length and $\sigma$ is the surface tension divided by fluid density.
While a solution to this equation may formally be written down in terms of Bessel and Lommel functions,
\begin{align}
 H(r) = -  C_p \dfrac{K_2^{2/5} \nu^{6/5}l_c^{2/5}}{\sigma} \left[ K_0\left(\dfrac{r}{l_c} \right) \int^{r/l_c}_0 \dfrac{I_0(s)}{s^{3/5}}~ds - I_0\left(\dfrac{r}{l_c} \right) \int^{r/l_c}_\infty \dfrac{K_0(s)}{s^{3/5}}~ds \right],
\label{eqn:lommel}
\end{align}
where the lower limits of integration are chosen to satisfy the conditions of regularity of $H$ at $r=0$ and $r\to\infty$.
For $r\gg l_c$, the gravitational force dominates over the surface tension, and the interface may be approximated by
\begin{align}
 H \approx C_p \dfrac{K_2^{2/5} \nu^{6/5} l_c^2}{ \sigma r^{8/5}}.
\end{align}
For $r\ll l_c$, the Laplace pressure from surface tension dominates over gravity, leading to the dominant balance
\begin{align}
\dfrac{1}{r} \left( r H_r \right)_r \approx  -  C_p \dfrac{K_2^{2/5} \nu^{6/5}}{ \sigma r^{8/5}} \quad \longrightarrow \quad H \approx  C_p \dfrac{K_2^{2/5} \nu^{6/5}}{ \sigma } \left(H_0 l_c^{2/5} - \dfrac{25r^{2/5}}{4} \right),
% H =  -  C_p \dfrac{25 K_2^{2/5} \nu^{6/5} r^{2/5}}{ 4\sigma } 
\end{align}
where $H_0 \approx 6.9484928934459429$ is a constant of integration determined from a numerical evaluation of the integral in \eqref{eqn:lommel}.
% The maximum displacement of the interface is finite and occurs at $r=0$, which is
% \begin{align}
%  H(r=0) =  C_p H_0 \dfrac{K_2^{2/5} \nu^{6/5} l_c^{2/5}}{ \sigma }.
% \end{align}
The slope of the interface must remain small for the validity of the approximations underlying the similarity solution. 
This criterion translates to
\begin{align}
 |H_r(r)| \approx C_p \dfrac{K_2^{2/5} \nu^{6/5}}{ \sigma } \dfrac{5}{2r^{3/5}} \ll 1 \quad \longrightarrow \quad r \gg \left(\dfrac{5C_p}{2}\right)^{5/3} \dfrac{K_2^{2/3} \nu^{2}}{ \sigma^{5/3} }.
\end{align}

\subsection{Dissolution-dominated surfactant transport}
\label{sec:solsim}

The boundary layer approximation, that so successfully describes the flow resulting in the case of surfactant transported in the adsorbed phase, fails for the case of dissolved surfactant.
The reason for this failure will be presented later.
% It is so because in the latter case, as the flow approaches the stagnant state away from the interface, the viscous and the inertial forces decay whilst remaining in balance with each other.
% Since the velocity decays as $1/r$, both the inertial and viscous terms in (\ref{eqn:rmom}-\ref{eqn:zmom}) decay as $1/r^2$.
% As a consequence, the far-field dominant balances of forces is different from what occurs in the conventional boundary layer theory.
However, a self-similar solution is still possible, as first demonstrated by \cite{Bratukhin1967}.
Here we present a comparison of their solution with our numerical results. 
In addition, we derive the asymptotic behavior of this solution in the physically relevant limit of $\Sc = \nu/D\gg 1$ and show that a universal self-similar profiles exist in the limit $K_3 \Sc/\nu^2 \gg 1$.

The solution is best presented in terms of spherical polar coordinates $(\rho, \theta)$ defined by $r = \rho \sin \theta$ and $z=-\rho \cos\theta$, with the corresponding velocity components $v$ and $q$ repectively. 
The self-similar ansatz in this case is
\begin{align}
 v = \nu \dfrac{\hat{f}(\theta)}{\rho}, \text{ } q = \nu \dfrac{\hat{g}(\theta)}{\rho}, \text{ and } c_3 = \dfrac{\nu}{\Gamma_3} \dfrac{\hat{h}(\theta)}{\rho},
\end{align}
in terms of the symbols $\hat{f}$, $\hat{g}$ and $\hat{h}$.
An exact solution of the Navier-Stokes equations is possible in this case, represented in terms of a single parameter $A$ as
\begin{subequations}
\begin{align}
 \hat{g}(\theta) = -2 \dfrac{\text{d}}{\text{d} \theta} \log F(\zeta), \quad \hat{f}(\theta) = - \dfrac{1}{\sin\theta} \dfrac{\text{d}}{\text{d}\theta} (\sin\theta ~ \hat{g}(\theta)), \quad \hat{h}(\theta) = \dfrac{\mu A (1+A)^\Sc}{F(\zeta)^{2\Sc}}, \label{eqn:bratukhinsimilarity} \\
 \text{ where }\zeta=1+\cos\theta, \quad  F(\zeta) = n_2 \zeta^{n_1} - n_1 \zeta^{n_2}, \quad \text{and} \quad n_{1,2} = \dfrac{1 \pm \sqrt{1+A}}{2}.
\end{align}
\end{subequations}
The parameter $A$ quantifies the strength of the Marangoni forcing (the surface velocity is $A\nu/2r$) and is related to the surfactant release rate by
\begin{align}
 \dfrac{K_3 \Sc}{\nu^2} = A (1+A)^{\Sc} \int_1^2 \left[ 4 \Sc^2 \zeta (2-\zeta) \dfrac{F'(\zeta)^2}{F(\zeta)^2} + 1 \right]~ \dfrac{\text{d} \zeta}{F(\zeta)^{2\Sc}}.
 \label{eqn:A}
\end{align}
% The asymptotics of this similarity solution for $\Sc \gg 1$ and $A\gg 1$ is the topic of this paper.

The results of the numerical solutions shown in Figure \ref{fig:soluble_raw_panels}, when rescaled according to the self-similar ansatz, collapse on the similarity solution \eqref{eqn:bratukhinsimilarity}, as shown in Figure \ref{fig:sol_bl_scaled_bratukhin}.
Note that, while the flow is self-similar, the profile shape is not universal, but varies with $A$, which depends on the strength of Marangoni stress as represented by ${K_3 \Sc}/{\nu^2}$ and $\Sc$. 
This dependence is shown in Figure \ref{fig:sol_bl_scaled_bratukhin}(d), for $\Sc$ from $\sqrt{10}$ to $10^{6}$, the range of $\Sc$ that encompasses many common chemical species.
This range of $\Sc$ corresponds to the asymptotic limit $\Sc \gg 1$.
This dependence from \eqref{eqn:A} has the asymptotic limits 
\begin{align}
A \sim {K_3 \Sc}/{\nu^2} \text{ for } A\ll 1 \qquad \text{and} \qquad A \sim \dfrac{2^{2/3} K_3^{2/3} \Sc^{1/3}}{\pi^{1/3} \nu^{4/3}}  \text{ for } A\gg 1.
\label{eqn:Ascaling}
\end{align}
In these two extremes, we expect a universal flow profile to emerge. 
In the case we are interested where $A \gg 1$, the solution \eqref{eqn:bratukhinsimilarity} simplifies in a region of thickness $\Delta n z = O(1)$ (transforming to cylindrical polar coordinates) to
\begin{align}
u(r,z) \approx \dfrac{A \nu}{2r} e^{z/r} \left\{ \dfrac{ \cosh\left( {\Delta n}^{-1} \right) }{ \cosh^2 \left( \hat{\xi} - {\Delta n}^{-1}\right) } - 2{\Delta n}^{-1} \dfrac{\sinh \left( \hat{\xi} \right) }{\cosh \left( \hat{\xi} - {\Delta n}^{-1} \right) } \right\} + O\left({\Delta n^{-2}} \right),
\label{eqn:soluniversal1}
\end{align}
where $\hat{\xi} = {\Delta n z}/{2r}$, $A$ and $\Delta n = \sqrt{1+A}$ are determined from \eqref{eqn:A} (or Figure \ref{fig:sol_bl_scaled_bratukhin}(d)).
In the limit of large $A$, the profile approaches
\begin{align}
 u(r,z) \approx \dfrac{A \nu}{2r} { \text{sech}^2 \left( \dfrac{z A^{1/2}}{2r} \right) } + O\left(\dfrac{1}{\Delta n} \right). \label{eqn:soluniversal2}
\end{align}
Varying $A$ in \eqref{eqn:soluniversal2} simply rescales the boundary layer thickness as $\delta=2r/A^{1/2}$ and the velocity magnitude as $A\nu/2r$, but does not change the leading-order shape of the profile.
In this sense, the profile is universal.
% However, note that the leading order approximation in \eqref{eqn:soluniversal2} fails to support any shear at $z=0$, and is therefore incapable of transmitting the Marangoni stress from the surfactant.

The flow profile in \eqref{eqn:soluniversal2} in fact describes a free radial jet forced by a steady point momentum source at the origin and a shear free boundary condition on the surface, as originally derived by \cite{Squire1955}.
% {\bf Expand the next two sentences in two paragraphs with equations etc. to support these claims.}
Consider the solution of \eqref{eqn:prandtlbleqn}, which decays as $1/r$ with the ansatz
\begin{align}
u(r,z) = \dfrac{a\nu}{r} \tilde{f}'(\xi), \quad w(r,z) = \dfrac{a^{3/2} \nu}{r}\left[\xi \tilde{f}'(\xi) - \tilde{f}(\xi) \right], \quad \xi = \dfrac{za^{1/2}}{r}, 
\end{align}
where $a$ is a constant to be determined as part of the solution.
The self-similar form $\tilde{f}(\xi)$ satisfies
\begin{align}
 \tilde{f}''' + \tilde{f}'^2 + \tilde{f}\tilde{f}'' = 0.
\end{align}
The solution to this equation that satisfies $w(r,z=0)=0$ and $u(r, z\to\-\infty)=0$ is
\begin{align}
\tilde{f}'(\xi) = \text{sech}^2 \left( \dfrac{\xi}{\sqrt{2}} \right), 
\label{eqn:solprandtl}
\end{align}
which is identical to the leading order of \eqref{eqn:soluniversal2} with $a=A/2$.
This boundary layer approximation applies in the limit $a \gg 1$, equivalent to the one under which \eqref{eqn:soluniversal2} was derived. 
However, note that \eqref{eqn:solprandtl} and \eqref{eqn:soluniversal2} are unable to satisfy the Marangoni stress condition at $\xi=0$.
Therefore, the value of $a$ remains undetermined at this order of boundary layer theory.
% A solution of Prandtl's boundary layer approximation that decays as $1/r$ simply yields this radial jet, and fails to match the Marangoni stress condition. 
Bratukhin and Maurin's solution is a perturbation of Squire's jet, where the Marangoni stress is supported by the next term of $O(\Delta n^{-1})$, and therefore the surface shear stress does not scale with dimensional parameters according to the leading order. 
Due to this structure of the flow, the simple scaling analysis of the type presented in \S\ref{subsec:numerical} fails to capture the dependence of dimensional pre-factors on the parameters. 
Using the next order correction in \eqref{eqn:soluniversal1} or the complete solution in \eqref{eqn:bratukhinsimilarity}, we see that the shear at the surface is $A\nu/r^2$ (weaker than the dimensionally expected $A \nu/\delta r$ as explained).
The shear stress on the surface $\mu A \nu /r^2$ thus balances the Marangoni stress, which scales as $\Gamma_3 c_3/r$, yielding the scale for surfactant concentration $c_3 \propto \mu A \nu / \Gamma_3 r$. 
Surfactant transport rate $q_3$ then scales as $r u c_3 \delta \Sc^{-1/2} = A^{3/2} \nu^2 \mu/\Gamma_3 \Sc^{1/2}$, where we have used that the surfactant boundary layer is thinner than the momentum boundary layer by a factor $\Sc^{1/2}$.
This final balance yields $A \propto (K_3^2 \Sc/\nu^4)^{1/3}$ in agreement with \eqref{eqn:Ascaling} and the results in Figure \ref{fig:sol_bl_scaled_bratukhin}(d).

\section{Discussion and conclusion}
\label{sec:conclusion}
Here we have presented the similarity solutions underlying the Marangoni stress driven flow due to a steady point source of surfactant, as it spreads either in an adsorbed or dissolved state.
The scope of this paper is limited to the case where the flow occurs in a thin layer close to the free surface, as characterized by $\delta \ll r$ for both cases, which in either case results in a universal velocity profile.

% The similarity solution for the case of the adsorbed surfactant is derived using Prandtl's boundary layer approximation to the Navier Stokes equations and presented in Figure \ref{fig:bl_scaled}, while in case of dissolved surfactant follows from the treatment of \cite{Bratukhin1967}, which is applicable whether or not $\delta \ll r$.
% For $K_3 \Sc/\nu^2 \gg 1$ and $\Sc \gg 1$, a universal boundary layer profile \eqref{eqn:soluniversal2} emerges.
% We analyze and elucidate the limit $\delta \ll r$, in which case a universal profile emerges.

Based on the analysis presented in this article, the following hydrodynamic signatures distinguish the two limits, expressed in terms of quantities that may be measured experimentally.
The power law exponent for the decay of surface radial velocity $u(r,z=0)$ is the first and obvious signature that distinguishes between the two limits.
The surface radial velocity decays proportional to  $r^{-3/5}$ in the adsorbtion-dominated case and to $r^{-1}$ in the dissolution-dominated case.
% Suppose that the velocity profile decay in the two cases is measured to be
% \begin{align}
%  u(r,z=0) &= C_a r^{-3/5} \quad \dots \quad \text{adsorption-dominated}, \\
%           &= C_d r^{-1}   \quad \dots \quad \text{dissolution-dominated}.
% \end{align}
The boundary layer scalings that collapse the depth-wise velocity profile constitute the second hydrodynamic signature that distinguish between the two limits.
Along with rescaling the velocity $u(r,z)$ with $u(r,z=0)$, the depth-wise coordinates need to be rescaled with factors of $\delta(r)=\sqrt{\nu r f'(0) /u(r,z=0)}$ for the profiles to collapse with $f(\xi)/f(0)$ in the adsorption-dominated case.
The appropriate depth-wise coordinate rescaling factor can be derived in the dissolution-dominated regime from \eqref{eqn:soluniversal2} to be $\sqrt{\nu r/ u(r,z=0)}$.
The third signature is the relation between the surface velocity and the surface shear stress at different radii.
In the adsorption-dominated case, this relation is
% \begin{align}
%  u(r,z=0) = \dfrac{K_2^{2/5} \nu^{1/5}}{r^{3/5}} f'(0) = \dfrac{b}{r^{3/5}}, \quad \text{and} \quad u_z(r,z=0) = \dfrac{K_2^{2/5} \nu^{1/5} }{r^{3/5} \delta(r)} f''(0).
% \end{align}
\begin{align}
 u_z(r, z=0) = \dfrac{u(r,z=0)}{\delta(r) } \dfrac{f''(0)}{f'(0)},
\end{align}
whereas for the dissolution-dominated case it is
\begin{align}
 u_z(r,z=0) = \dfrac{2u(r,z=0)}{r}.
\end{align}
Depending on the available experimental accuracy, successively stricter comparison of the measured velocity profile with the theory presented in this manuscript can be made using these three signatures.

% The nature of the profile is 
Our results rationalize experimental observations by \cite{Mandre2017a} of the power-law decay and boundary layer structure of Marangoni-driven flow.
The agreement between the experimental measurements and our theory suggests that the more general surfactant dynamics may under certain circumstances be reduced to simple models akin to Marangoni and Gibbs elasticity of surfactant-laden liquid interfaces.
The solutions developed here could also provide insight into the surfactant dynamics based on flow velocimetry in experimental systems, such as by \cite{Roche2014} and \cite{LeRoux2016}.
The precise criteria under which simplification is possible and conditions for transition are left to be undertaken in the future.

% \section{Conclusion}
% \section{}
% Figure \ref{fig:python3program} shows a simple Python3 program for computing the solution of \eqref{eqn:selfsimilarode} and \eqref{eqn:selfsimilarbc}. 
% 
% \begin{figure}
% \begin{centering}
% \begin{verbatim}
% import numpy as np
% 
% def fun(t, y):
%     f = y[0]
%     fp = y[1]
%     fpp = y[2]
%     fppp = - 0.6*fp**2 - 1.2*f*fpp 
%     return np.array([fp, fpp, fppp])
%  
% t = 0
% N = 600000
% dt = -1e-4
% T = np.linspace(0, dt*N, N+1)
% fpp0 = 0.402287361293201
% y = np.array([0, 0.4/fpp0, fpp0])
% Y = np.zeros((N+1,3));
% 
% ii = 0
% for ii in range(N+1):
%     Y[ii,:] = y[np.newaxis, :]
%     dy1 = dt*fun(t, y)
%     dy2 = dt*fun(t+dt/2, y+dy1/2)
%     dy3 = dt*fun(t+dt/2, y+dy2/2)
%     dy4 = dt*fun(t+dt, y+dy3)
%     y = y + (dy1+2*dy2+2*dy3+dy4)/6
%     t = t + dt
% 
% \end{verbatim}
% \end{centering}
% \caption{Python3 program to solve the initial value problem for $f(\xi)$. 
% The program converts the third order ordinary differential equation \eqref{eqn:selfsimilarode} to three first order equations for the variables \texttt{y[0]}, \texttt{y[1]}, and \texttt{y[2]}, and integrates them using the standard fourth order Runge-Kutta method.}
% \label{fig:python3program}
% \end{figure}

\bibliographystyle{jfm}

 % ../../bib/cboat}
\end{document}